Rotational Splitting of Pulsation Modes
Suggested Running Head: Rotational Splitting

Robert G. Deupree and Wilfried Beslin
Institute for Computational Astrophysics and Department of Astronomy and Physics,
Saint Mary's University, Halifax, NS B3H 3C3, Canada bdeupree@ap.smu.ca

ABSTRACT

Mode splittings produced by uniform rotation and a particular form of differential rotation are computed for two-dimensional rotating 10 $M_o$ ZAMS stellar models. The change in the character of the mode splitting is traced as a function of uniform rotation rate, and it is found that only relatively slow rotation rates are required before the mode splitting becomes asymmetric about the azimuthally symmetric (m=0) mode. Increased rotation produces a progressively altered pattern of the individual modes with respect to each other. Large mode splittings begin to overlap with the mode splittings produced by different radial and latitudinal modes at relatively low rotation rates. The mode splitting pattern for the differentially rotating stars we model is different than that for uniformly rotating stars, making the mode splitting a possible discriminant of the internal angular momentum distribution if one assumes the formidable challenge of mode identification can be overcome.

1.  INTRODUCTION

One particular area of interest in asteroseismology is the possibility of determining something about the internal angular momentum distribution of individual stars, including those which rotate rapidly. There are rapidly rotating stars which exhibit nonradial stellar oscillations (e.g., Cameron, et al. 2008). Many of these oscillations are g modes, but there are also β Cephei variables which rotate fairly rapidly and exhibit p mode oscillations (Balona, Dziembowski & Pamyatnykh 1997). While rotation alters the frequencies of the individual axisymmetric modes (e.g., Saio 1981; Lignières, Rieutord & Reese 2006; Lovekin & Deupree 2008), it also lifts the degeneracy in frequency for the nonaxisymmetric modes. This mode splitting has been observed in a variety of stars (e.g., Zwintz, et al. 2009; Balona 2002), but certainly including β Cephei variables (e.g., Aerts et al. 2004; Handler, et al. 2004; Jerzykiewicz, et al. 2005). Most of the cases identified have been of stars rotating sufficiently slowly that the mode splitting is comparatively small and easy to identify. More rapid rotation has been expected to lead to problems with mode identification at least because the rotational mode splitting becomes nonuniform and because the relatively straightforward classification of modes by a single spherical harmonic becomes invalid (e.g., Lignières, Rieutord & Reese 2006; Reese, Lignières & Rieutord 2006, 2008; Lovekin & Deupree 2008; Goupil, et al. 2005; Breger, Lenz, & Pamyatnykh 2009).

For sufficiently small rotation rates, the mode splitting in the inertial frame is linearly proportional to mΩ, where Ω is the rotation rate and m is the azimuthal quantum number. This linear relationship is produced predominantly by the transformation from the rotating frame of the star to the inertial frame. Once the rotation rate becomes sufficiently large, both the centrifugal and higher order Coriolis force terms become important and require the addition of a term proportional at the lowest order to $\Omega^2$. One of the objectives of this work is to examine at what rotation rate the mode splitting is no longer linear for the low order p modes we shall consider. The increase of the mode splitting with rotation suggests that the mode splitting could become quite large as the rotation rate becomes large, assuming the mode splitting is not decreased by the higher order effects. Large mode splittings would present a problem for mode identification because the frequency splitting between adjacent values of m for a given n and $\ell$ could become the same size or even larger as the frequency separation between adjacent values of $\ell$ for a given n and m, or even the frequency separation between adjacent values of n for given $\ell$ and m. In fact, calculations of nonaxisymmetric modes for uniformly rotating, uniform density models by Espinosa, et al. (2004) do show that this overlap in the size of the frequency separation does occur. We wish to explore this phenomenon with more realistic stellar models and less restrictive linearized pulsation calculations than these authors used.

There are two steps required to compute the pulsation frequencies of rapidly rotating stars. The first is to compute the rotating models themselves. For this we use the 2.5 D, fully implicit stellar hydrodynamics and evolution code developed by Deupree (1990, 1995). This code solves six conservation laws (mass, three components of momentum, energy, and hydrogen abundance) along with Poisson's equation (and the usual subsidiary relations for the equation of state, nuclear reaction rates, and radiative opacity). The independent variables are the fractional surface equatorial radius and the colatitude. The models computed here are 10 $M_o$, ZAMS models for which the rotation rate is imposed and the equations of hydrostatic and thermal balance are solved implicitly to obtain the structure. The conservation laws do not make any assumptions about the rotation rate except that the model must be azimuthally symmetric, although the surface is located by assuming that the surface is an equipotential.

Once the rotating models have been obtained, we compute the pulsational frequencies using the linear, adiabatic, nonradial pulsation code developed by Clement (1998) and modified by Lovekin, Deupree, & Clement (2009) to include differential rotation. This code allows the eigenfunctions to be expressed as a sum of N spherical harmonics whose coefficients are determined by the radial integration of the linearized equations at N specific latitudes. We will take N = 6, which Lovekin & Deupree (2008) have found to yield accurate eigenfrequencies for the rotation rates examined. For sufficiently rapidly rotating models, the latitudinal spherical harmonic quantum number, $\ell$, no longer has a unique value for the resulting eigenfunction. This generates a nomenclature issue which we resolve by characterizing the mode by $\ell_0$, the value of $\ell$ associated with the mode in the nonrotating model to which the current mode can be traced. Such tracing becomes progressively more difficult for more rapidly rotating models.

We shall restrict our concern to low order p modes, particularly those with radial nodes (designated by n) of 1 - 3; latitudinal quantum numbers ($\ell_0$) of 1 – 3, and azimuthal quantum numbers (m) ranging from - $\ell_0$ to $\ell_0$. Our restriction to low radial quantum numbers suggests that the asymptotic approach described by Reese, et al. (2009) may not be appropriate here. It should be noted that the range of m may not cover all possible values for a rotating model in which the dominant value of ℓ is larger than $\ell_0$. Both uniform and differentially rotating models are considered. The surface equatorial velocities for the uniformly rotating models range from 0 to 360 km s$^{-1}$, which may be compared with the critical rotation velocity of about 600 km s$^{-1}$. We will discuss the results in terms of the surface equatorial velocity for simplicity, but will present the more relevant quantity, $\Omega/\Omega_{crit}$, where it might be useful. Here $\Omega_{crit}$ is the critical rotation rate. The differentially rotating models are computed with a rotation law which is a generalization of that given by Jackson, MacGregor, and Skumanich (2005) in their attempt to match the observed oblateness then attributed to the surface of Achernar (Domiciano de Souza, et al. 2004):

$$\Omega(\varpi) = \frac{\Omega_0}{1 + (a\varpi)^\beta}$$

where $\Omega$ is the rotation rate, $\varpi$ is the distance from the rotation axis in units of the surface equatorial radius, and $\Omega_0$, a, and β are constants. The value of β must be less than 2 for stability, the constant a determines the distance from the rotational axis at which the transition from increasing rotation rate to flat rotation rate very close to the axis is made, and the constant $\Omega_0$ determines the overall amount of rotation once a and β are chosen. We have computed models with β = 0.2, 0.6, 1.0, 1.4, and 1.8 for a surface equatorial velocity of 120 km s$^{-1}$, and β = 0.2, 0.4, 0.6, 0.8, and 1.0 for a surface equatorial velocity of 240 km s$^{-1}$.

The axisymmetric modes for these models have been presented by Lovekin, Deupree & Clement (2009). Here we shall focus on the nonaxisymmetric modes, in particular on the rotational splitting of the modes.

## 2. ROTATIONAL SPLITTING
### *2.1 Uniform Rotation*

We show the pulsation frequencies as a function of surface equatorial velocity for uniformly rotating models in Figure 1. The frequencies in this and subsequent figures are presented in units of $(4\pi G)^{1/2} = 9.157 \times 10^{-4}$, and are inertial frame frequencies. The modes in Figure 1 have n = 1 and $\ell_0$ = 3. The shape of the curves is reminiscent of that of Espinosa, et al. (2004) for rotating uniform density objects. We note Clement (1998) and Espinosa, et al. (2004) have defined the direction of positive m in opposite directions, as may be seen by comparing Clement's equation (5) with Espinosa, et al.'s equation (1). This explains why our results show higher frequencies for negative m and Espinosa, et al. have higher frequencies for positive m.

The variation in frequency versus rotation shown in this figure results from several sources: the linear variation proportional to m which arises from the transformation from the rotating to the inertial frame, the (much smaller) linear variation proportional to m arising from the Coriolis force produced by the pulsation perturbations on the originally rotating structure, the frequency shifts produced in the static structure which is changed by the centrifugal force, and the other effects of the perturbations acting on this structure. We will focus first on the rotation rate at which first order perturbation theory begins to break down for these low order p modes, and we will do that by focusing on the frequency differences of the nonazimuthal modes, i.e.,

$$D = \omega_{n,\ell_0}(m) - \omega_{n,\ell_0}(m=0)$$

If the rotation rate is sufficiently small, the mode splitting term linearly proportional to m will dominate. Under this condition, the spacing should be uniform because the frequency shift is linearly proportional to m, and the magnitude of the slope of the spacing could be computed by

$$E = \frac{|D|}{|m|}$$

for any nonzero value of m. If the mode splitting is linear, E will have the same value for all nonzero values of m. We present the results of this calculation for all nonzero m values of a mode for slowly rotating models in Figure 2. From this figure it is clear that only a modest amount of rotation is required before assuming a linear relationship for all mode separations breaks down. There is a noticeable difference between the separation for positive values of m and for negative values of m for the comparatively slow rotation speed of 30 km s$^{-1}$ ($\Omega/\Omega_{crit} \approx 0.076$). By 50 km s$^{-1}$ ($\Omega/\Omega_{crit} \approx 0.126$) there is a clear separation for all values of m. The picture is more complex at higher rotation rates as the centrifugal force and higher order Coriolis terms become important.

The net result of this is that the mode splitting becomes non uniform at relatively slow rotation rates and dramatically so as the rotation rate increases. We present the frequency separation for the different values of m in comparison to the m = 0 mode for various rotational velocities in Figure 3. Generally speaking, the |m| = 1 modes show more deviation from a straight line at higher rotation in plots like Figure 3. There are some hints that the |m| = 2 modes show more deviation than the |m| = 3 modes, but neither is as compelling as for the |m| = 1 modes.

One detail worth mentioning from Figure 1 is that the mode splitting for sufficiently high rotation becomes larger than the frequency change between the nonrotating and rotating models for the axisymmetric mode. This is true for all values of n and $\ell_0$ considered here. This fact and the separation of modes in the nonrotating case have significant consequences for the mode distribution. The effects may be seen in Figure 4, a one dimensional plot of the frequencies for all the modes computed here at several rotation velocities. Figure 4a shows the frequencies at 30 km s$^{-1}$, and it is seen that the frequency pattern conforms to n spacings (i.e., the spacing of frequencies for adjacent values of n) being larger than $\ell_0$ spacings, which are in turn larger than m spacings. This pattern is already disrupted by a rotation rate of 90 km s$^{-1}$ ($\Omega/\Omega_{crit} \approx$ 0.226) in Figure 4b, where we already have overlap between different $\ell_0$ modes and are just beginning to have overlap for different n modes. Looking at the last two parts of Figure 4, we see that gaps are produced and removed as the rotation rate increases, with no real correlation to gaps at other rotation rates. In Figure 4d only the gaps near the center of the plot can be considered gaps, as other modes we did not calculate would produce frequencies in the high and low frequency parts of this diagram. This complicated frequency spectrum suggests that an independent, observational determination of at least some of n, $\ell_0$, and m may be essential to correlate the frequencies to interior structure properties for sufficiently rapidly rotating stars. We shall explore this further in the next section.

## 2.2 Differential Rotation

Lovekin, Deupree & Clement (2009) computed frequencies for both uniformly and differentially rotating models. The differentially rotating models utilized the rotation law given in equation (1) with the parameters mentioned in the introduction. One result of this work was that the frequencies depended in a relatively significant way on the amount of rotation the model possessed, but did not depend on the distribution of rotation within the model (at least for this particular rotation law). For example, differentially rotating models produced the same trends in frequency as uniformly rotating models with slightly higher surface rotation velocities. This led us to examine if there was any more information about the interior angular momentum distribution in the nonaxisymmetric modes.

Fortunately, there is. This was determined by comparing the mode splitting for a differentially rotating model with the mode splitting for a uniformly rotating model. The uniformly rotating model used for comparison was determined by forcing the m = - $\ell_0$ mode to have the same splitting from the m = 0 mode for the same values of n and $\ell_0$. This comparison uniformly rotating model had a slightly higher surface equatorial velocity than did the differentially rotating model. We compare the m splittings for an interpolated model uniformly rotating at 165 km s$^{-1}$ with one rotating with a surface rotational velocity of 120 km s$^{-1}$ and β = 1.4 in Figure 5. This figure has n = 2, $\ell_0$ = 3. There is little dependence on n and a slight dependence on $\ell_0$. Nor is there any significant dependence of the difference in splitting between uniformly and differentially rotating models with increasing surface equatorial velocity for a given value of β. Of course, decreasing β decreases the difference in splitting, but differences remain at approximately 0.01 in our scaled frequency space (this translates into approximately nine μHz for this particular model) for β = 0.6.

While this distinction between m splittings for uniformly and differentially rotating models is desirable, one must be able to unravel the interweaving of modes with different values of n and $\ell_0$ amongst the various m modes. We have examined this in several ways in the next section.

## 3. MODE IDENTIFICATION

A quick glance at the frequency distributions for the more rapidly rotating members of Figure 4 indicates that our ability to identify modes and thus assign the proper mode spacing is in jeopardy. We have examined several ways to attempt to recover this information. Before going into these, we present a few details that appear to be true for all rotation rates above 90 km s$^{-1}$ (this restriction arises from the small variation amongst the different separations at lower rotation speeds):

- the smallest spacing magnitude is always between m = -$\ell_0$ and -$\ell_0$ + 1 for any $\ell_0$,
- the smallest spacing magnitude decreases as n increases for any $\ell_0$,
- the smallest spacing magnitude in our sample is always for the mode n = 3, $\ell_0$ = 1 and m = -1 and 0,
- spacings for a given n and $\ell_0$ show considerably less variation with $\Delta$m = 2 than with $\Delta$m = 1,
- for $\ell_0$ = 1, the magnitude of the frequency separation between m = 1 and m = -1 modes is nearly independent of n,
- for $\ell_0$ = 2, there is an increase in the magnitude of the frequency separation between the m = 2 and m = -2 modes with increasing n (with one counterexample),
- for $\ell_0$ = 3, there is an increase in the frequency separation between the m = 3 and m = -3 modes with increasing n,
- for a given n, the magnitude of the smallest separation for a given $\ell_0$ increases as $\ell_0$ increases (this is not quite true for n = 1, as the smallest separation for $\ell_0$ = 2 is less than that for $\ell_0$ = 1), and
- for $\ell_0$ = 3, the magnitude of the $\Delta$m = 4 separations is ordered in the following way: $|f_{m=1} - f_{m=-3}| < |f_{m=2} - f_{m=-2}| < |f_{m=-1} - f_{m=3}|$ for a given n. At a given rotation speed the overlap between any of one of these three separations with any of the others is quite small; however, the magnitude of all these frequency separations depends on the rotation rate. We note in passing that these results are based solely on the rotational sequence for this one model and have only been deduced for these low order p modes with relatively low $\ell_0$. Future work will determine whether these findings have wider applicability.

The first approach attempted to identify the modes by computing the Fourier transform of the frequency spectrum (e.g., Chaplin, et al. 2008). The scaled frequency interval between 0.8 and 1.2 was subdivided into intervals of 0.0001, this covering the number of digits retained for each computed frequency. If there was no actual frequency in this interval, we assigned a value of zero for its weight; otherwise a value of one was assigned. This distribution of the weight as a function of scaled frequency provided the input for the Fourier transform. Thus, our input could be taken to represent a very clean frequency power spectrum. When the rotation rate was low, the power spectrum of this Fourier transform had the expected behavior - the power distribution replicated itself around integer multiples of the "time interval" corresponding to the frequency spacing in m. As the rotation rate is increased, the distributions move to lower time intervals (corresponding to higher frequency spacings in m) and begin to overlap and spread out. Once the rotation rate is sufficiently large, there is nothing remarkable about range in the power spectrum which corresponds to the range in m spacing. We show one of these power distributions in Figure 6, with the vertical lines marking the intervals corresponding to the range of frequency spacings in m for the model rotating uniformly with a surface equatorial velocity of 180 km s$^{-1}$ ($\Omega$ /$\Omega_{crit}$ ≈ 0.444). There are no clear peaks in the power distribution that stand out with respect to those peaks outside this time interval range. We conclude that Fourier transforms alone will not lead to a successful unraveling of the complex nature of these frequency spectra.

The magnitude of the problem can perhaps be demonstrated in Figure 7. This is a plot of three histograms for a model rotating at a surface equatorial velocity of 180 km s$^{-1}$. One of these plots is the actual spacings for $\Delta m = 1$. Another histogram is the separations between two adjacent frequencies in the distribution. It is clear that these two distributions are quite different. This again just shows that, once the model is rotating sufficiently rapidly, that the $\Delta m = 1$ separations become large. We had hoped that we could use this fact by deleting those separations which are very small and replacing them by the separation with the next frequency (i.e., not the initially next frequency, but the frequency after that). This is the third histogram, and one can see that this does not come very close to the actual $\Delta m = 1$ separations in the first histogram. While we could continue the process, unfortunately one must make a judgment about what separation can no longer be deleted. In the end we decided this was not a very fruitful pursuit.

One final attempt arose using the information contained in the trends discussed above. We wanted to see if these patterns could be used to isolate specific modes. The example we chose was the m = 3 modes for the specific surface equatorial velocity of 150 km s$^{-1}$ ($\Omega/\Omega_{crit} \approx 0.373$). This rotation rate was chosen because it is sufficiently large that there is overlap among the various n and $\ell_0$ values, but not so large that many modes we did not compute should be included in the frequency interval. Clearly, with m = 3, we have $\ell_0 = 3$. We began by computing the frequency separation between the m = 3 mode and the six other m values for a given n. This gives us three values, one for each n, for the frequency separation between the m = 3 mode and any one other value of m. We take the range of variation of these three variables as the allowable frequency separation window, giving us a total of six windows. For every frequency in the spectrum, we make the assumption that it is an m = 3 mode and test to see if there is an actual frequency in each of the six windows. There are 34 frequencies to which we can apply the method (the remaining eleven have windows which extend outside the frequency range of the calculations and were thus not considered). Of these, twelve, including the three correct ones, met the condition that there was a computed frequency present in each of the six windows. We have looked at a few more criteria based on the patterns, but the result is always that there are several false positives.

It should be emphasized that we made this test case about as benign as possible. There are no questionable frequencies, and we assumed that we knew exactly what the frequency boundaries of each window were. We also assumed that all frequencies for a given n and $\ell_0$ would be present. The window boundaries do increase as the rotation rate increases, so we cannot assume we know exactly what these boundaries are for a given set of observations. We repeated the exercise applying the window boundaries for a surface equatorial velocity of 240 km s$^{-1}$ to the 150 km s$^{-1}$ frequency data set and found that we still had three frequencies which contained computed frequencies in all six windows. None of these were any of the actual m = 3 frequencies.

These results indicate that it is going to be very difficult to identify the pulsation modes for a sufficiently rapidly rotating star merely using the observed frequencies. Of course, we have by no means exhausted possible identification mechanisms, but we believe we have done enough to demonstrate that the problem is nontrivial.

## 4. DISCUSSION

We have computed nonaxisymmetric modes for low order p modes of 10 $M_o$ ZAMS models with various amounts of rotation. Both uniform rotation and rotation laws in which the rotation rate increases with decreasing distance from the rotation axis have been considered. At low uniform rotation rates the frequency rotational mode splitting is essentially the same for all modes, as expected. Sizeable departures from this uniformity appear at rather low rotation rates (surface equatorial velocity $\approx 50$ km s$^{-1}$ or $\Omega / \Omega_{crit} \approx 0.126$). The rotation rate is still relatively small (surface equatorial velocity $\approx 90$ km s$^{-1}$ or $\Omega / \Omega_{crit} \approx 0.226$) when the rotational splitting becomes sufficiently large so that modes of different n or $\ell_0$ begin to overlap. This overlap greatly complicates mode identification, and we have not been able to find a method which solves this problem using only the frequency spectrum.

A further complication is that rotational mode splitting is fairly sensitive to differential rotation, at least for the differential rotation law we used. This sensitivity is greater than we have observed for the axisymmetric modes and may be the best hope for constraining the interior angular momentum distribution using asteroseismology in the near future, assuming the problem of mode identification can be solved.

This research was supported through a Discovery grant by the National Science and Engineering Research Council of Canada. Computational facilities were provided by grants from the Canada Foundation for Innovation and the Nova Scotia Research Innovation Trust.

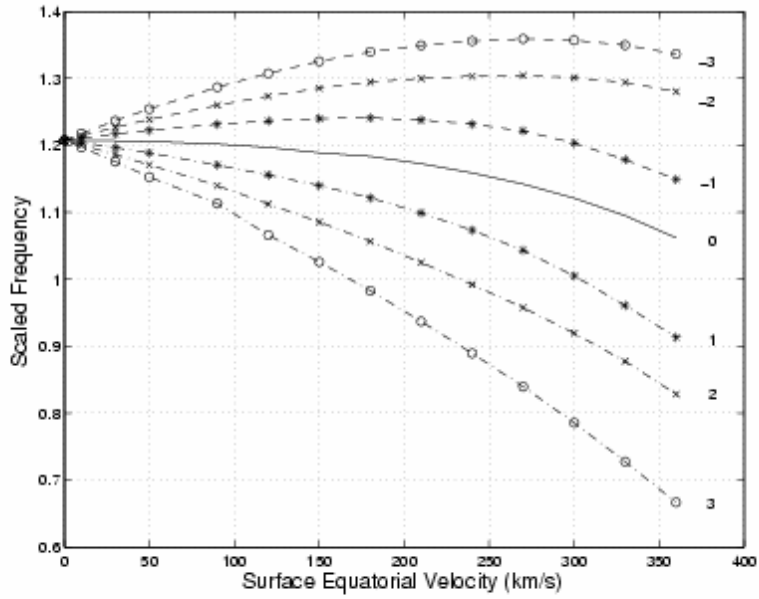

Figure 1. Scaled frequency versus surface equatorial rotational velocity for 10 M$_\circ$ ZAMS uniformly rotating models. All frequencies in this and subsequent figures are scaled by $(4\pi G)^{1/2}$. The modes have n = 1 and $\ell_0$ = 3. The values of the azimuthal mode identifier, m, are given at the right.

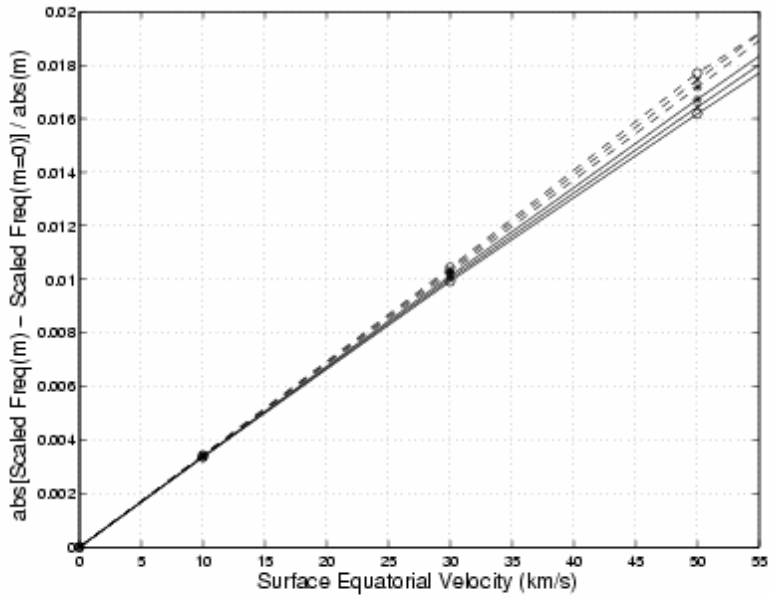

Figure 2. Plot of E(m) versus surface equatorial rotational velocity for selected 10 $M_\odot$ ZAMS uniformly rotating models. Solid lines denote negative values of m, dashed lines positive values. Circles, crosses, and asterisks denote absolute values of m of 3, 2, and 1, respectively. Uniform spacing in rotational splitting would result in a single, straight line.

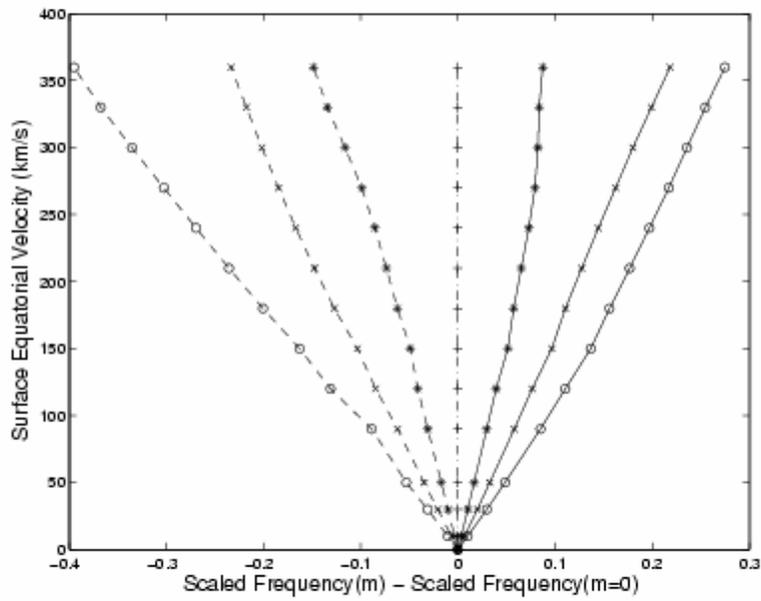

Figure 3. Plot of surface equatorial rotational velocity versus D(m). Solid lines denote negative values of m, dashed lines positive values. Circles, crosses, asterisks, and pluses denote absolute values of m of 3, 2, 1, and 0, respectively.

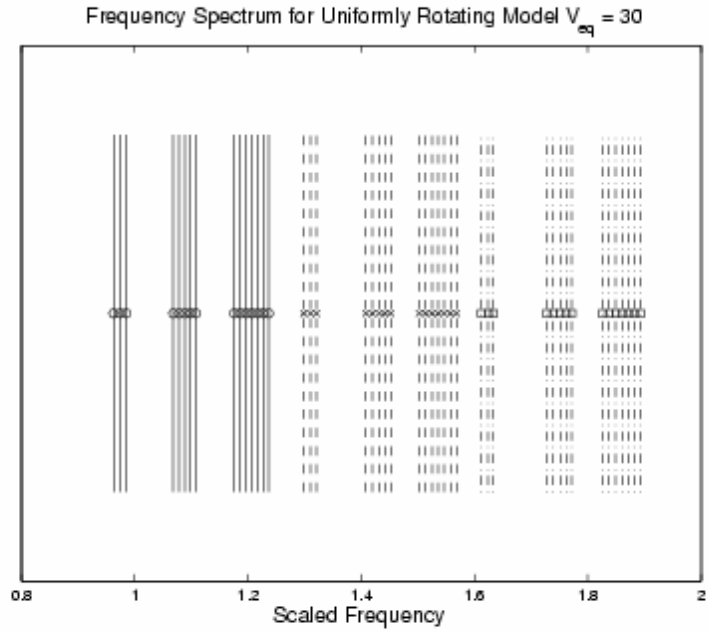

Figure 4a. Computed frequencies for a 10 M$_\circ$ ZAMS model rotating uniformly with a surface equatorial velocity of 30 km s$^{-1}$ ($\Omega/\Omega_{crit} \approx 0.076$). All modes computed are displayed. Solid lines, dashed lines, and dash-dot lines denote modes with n = 1, 2, and 3, respectively. Note that the separation between adjacent values of n is much larger than frequency separations for adjacent values of $\ell_0$ for the same n, which are in turn much larger than separations for adjacent values of m for the same n and $\ell_0$. This pattern greatly aids in mode identification.

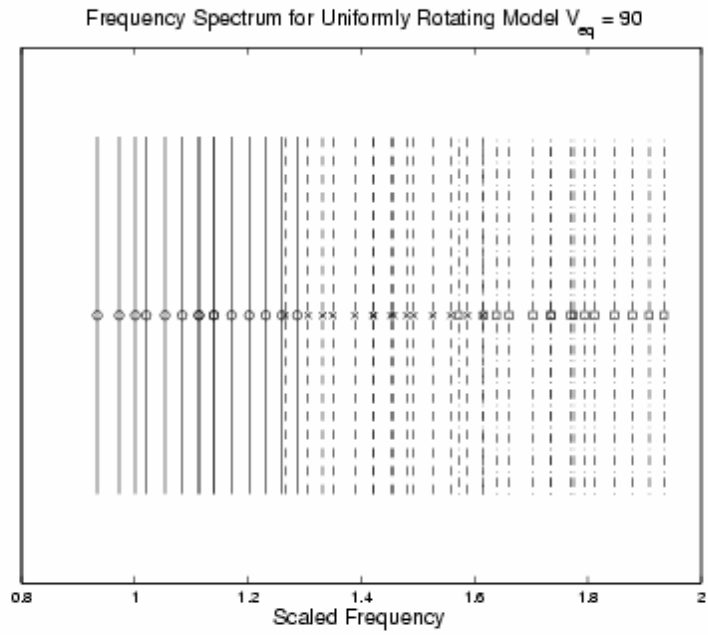

Figure 4b. Computed frequencies for a 10 M$_\circ$ ZAMS model rotating uniformly with a surface equatorial velocity of 90 km s$^{-1}$ ($\Omega/\Omega_{crit} \approx 0.226$). Symbols are the same as in Figure 4a. Note that the frequency separation in m is now approximately the same as the frequency separation in $\ell_0$, and that the frequencies of different values of n overlap.

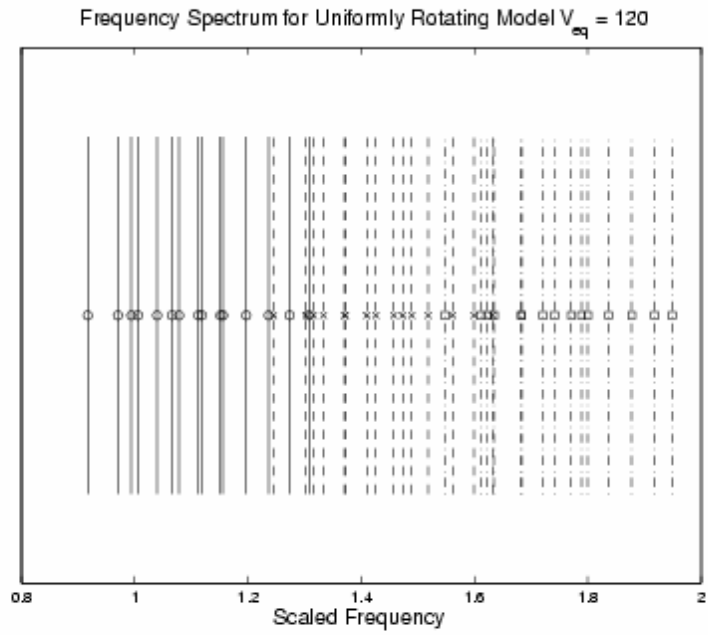

Figure 4c. Computed frequencies for a 10 M$_\circ$ ZAMS model rotating uniformly with a surface equatorial velocity of 120 km s$^{-1}$ ($\Omega$ /$\Omega_{crit}$ ≈ 0.300). Symbols are the same as in Figure 4a. There is overlap of the different separations, creating some new frequency gaps and closing some other ones.

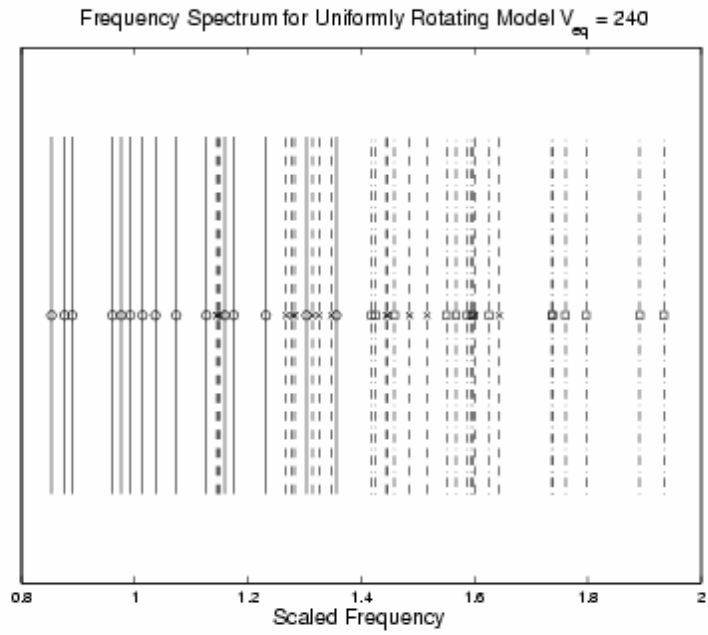

Figure 4d. Computed frequencies for a 10 M$_\circ$ ZAMS model rotating uniformly with a surface equatorial velocity of 240 km s$^{-1}$ ($\Omega/\Omega_{crit} \approx 0.578$). Symbols are the same as in Figure 4a. The relative locations of gaps and clusters of frequencies continue to change. Caution should be taken in examining frequencies near the low and high frequency boundaries of this plot because frequencies we did not calculate would be present in these regions if included.

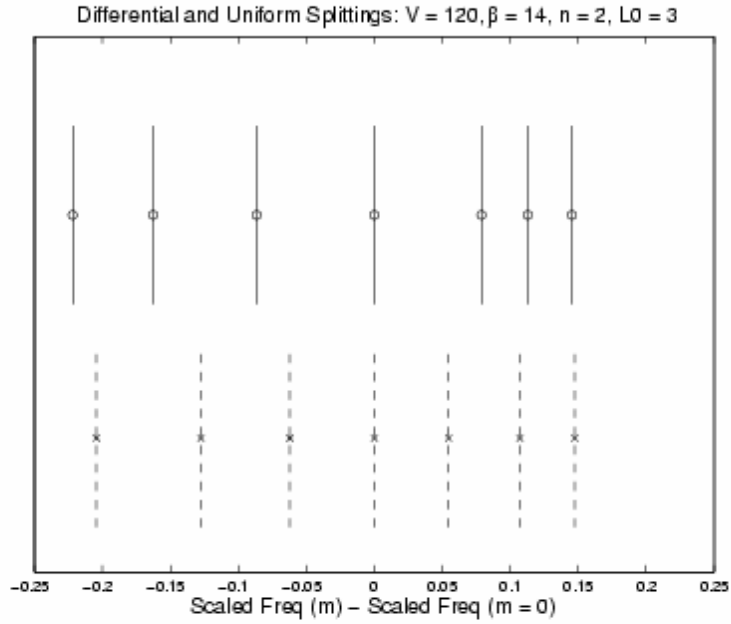

Figure 5. Comparison of m mode frequencies with n = 2, $\ell_0$ = 3 for an interpolated model rotating uniformly at 165 km s$^{-1}$ (dashed lines) with a differentially rotating model with β = 1.4 and a surface equatorial velocity of 120 km s$^{-1}$ (solid lines). These two were compared because the largest positive separation is nearly the same in the two cases. Note that the relative distribution of the nonazimuthal modes is significantly different for the uniformly rotating model from that of the differentially rotating model. This has the potential to be a useful discriminant for information about the interior angular momentum distribution.

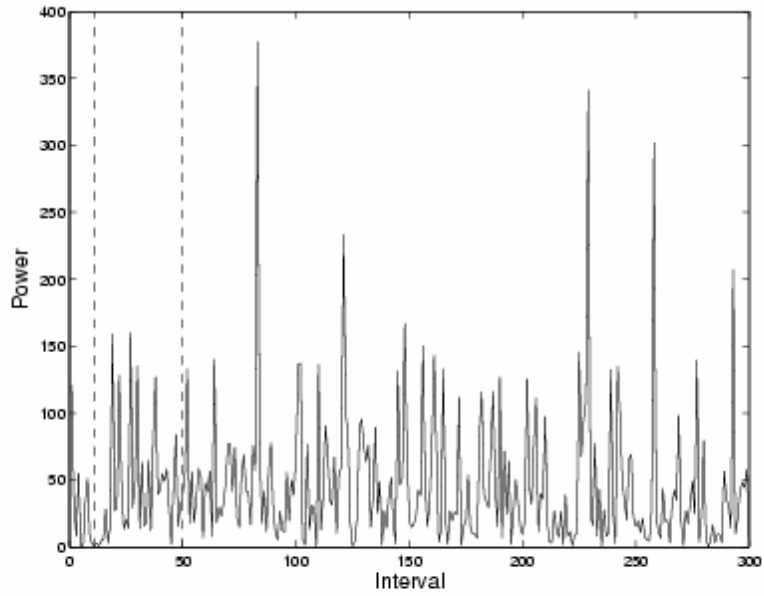

Figure 6. Power spectrum of the Fourier transform of the frequency spectrum for a model rotating uniformly at 180 km s$^{-1}$ ($\Omega/\Omega_{crit} \approx 0.444$). The vertical dashed lines correspond to the range of intervals that correspond to the range of frequency separations of adjacent nonaxisymmetric modes. Note that there are no distinguishing characteristics of the power spectrum in this interval and outside of it.

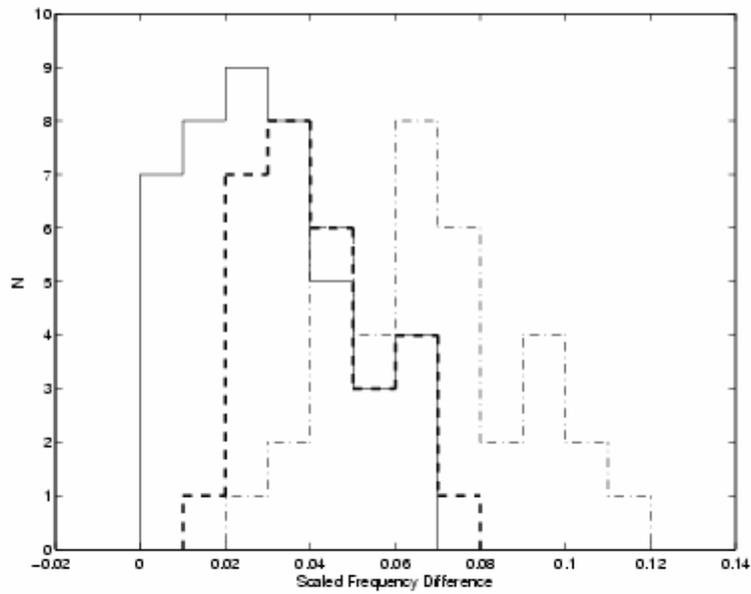

Figure 7. Histogram of three distributions of frequency separations. The solid curve denotes the distribution of the separation of adjacent frequencies in the computed frequency spectrum. The heavy dashed curve is the distribution that results when frequency separations which are too small to be related to m spacings are arbitrarily removed. The dashed-dot distribution is the actual separation between adjacent m mode frequencies.